# Strain-dependent resistance and giant gauge factor in monolayer WSe$_2$

Mao-Sen Qin(秦茂森), Xing-Guo Ye(叶兴国), Peng-Fei Zhu(朱鹏飞), Wen-Zheng Xu(徐文正), Jing Liang(梁晶), Kaihui Liu(刘开辉), Zhi-Min Liao(廖志敏)*

*State Key Laboratory for Mesoscopic Physics and Frontiers Science Center for Nano-optoelectronics, School of Physics, Peking University, Beijing 100871, China*
\* liaozm@pku.edu.cn

**We report the strong dependence of resistance on uniaxial strain in monolayer WSe$_2$ at various temperatures, where the gauge factor can reach as large as 2400. The observation of strain-dependent resistance and giant gauge factor is attributed to the emergence of nonzero Berry curvature dipole. Upon increasing strain, Berry curvature dipole can generate net orbital magnetization, which would introduce additional magnetic scattering, decreasing the mobility and thus conductivity. Our work demonstrates the strain engineering of Berry curvature and thus the transport properties, making monolayer WSe$_2$ potential for the application in the high-performance flexible and transparent electronics.**

## 1. Introduction

Two-dimensional (2D) van der Waals (vdW) layered transition metal dichalcogenides (TMDCs), such as MoS$_2$ and WSe$_2$, have demonstrated potential applications in the next-generation electronics owing to their remarkable electrical, optical, and mechanical properties[1-5]. Unlike the bulk semiconducting materials that are usually brittle, it is reported that TMDCs can exceptionally sustain in-plane strain as high as 11% in their monolayer counterparts[6], promising for the high-performance flexible and transparent electronics[7-10]. Based on their high flexibility, extensive theoretical and experimental studies have recently been motivated by the interests for strain engineering[11-28]. It is found strain can significantly affect the physical properties of TMDCs. Giant valley shift[21] and bandgap engineering[11,12,14,16] can be achieved by introducing uniaxial strain in monolayer TMDCs. Importantly, strain is also proposed to manipulate the Berry curvature[21,28], which is generally important in the topological transport. The uniaxial strain can break the $C_{3v}$ symmetry and thus induce asymmetrically distributed Berry curvature in a single valley, i.e., nonzero Berry curvature dipole in monolayer TMDCs, leading to the nonlinear Hall effect[28]. Furthermore, the strain is known to induce pseudo-magnetic field in 2D materials, which has been predicted and experimentally found to generate zero-field quantum Hall effect by forming Landau levels[17,25,26]. Additionally, for the TMDCs with orthorhombic phase, such as the Weyl semimetal $MoTe_2$, it is reported that the strain can effectively tune the Weyl points, where a small strain can annihilate two pairs of Weyl points and induce the rest pairs of Weyl points changing from type-II to type-I Weyl points[29]. Interestingly, in the twisted TMDCs proposed recently, the formation of moiré pattern can be actually regarded as strain engineering as well, which can result in various exotic phenomena including spin-liquid states, emergent magnetic order, and chiral superconductors[30]. Due to their high flexibility and strain-tunability, TMDCs provide an ideal platform to study the strain engineering of transport properties.

In this work, we investigate the strain-dependent resistance of the ionic gated monolayer $WSe_2$, where giant gauge factor at various temperature ranges is obtained. We produce uniaxial strain by the inverse piezoelectric effect of (1-x)Pb $(Mg_{1/3}Nb_{2/3})O_3$–$x$[$PbTiO_3$] (PMN-PT) with very high piezoelectric coefficient[31-35]. Under uniaxial tensile strain, the sheet resistance of $WSe_2$ dramatically changes. The extracted gauge factor is approximately 1100 at 140 K and is further enhanced to ~ 2400 at 2 K. Detailed discussions are carried out and this giant gauge factor is attributed to the strain-modulated mobility, which is suppressed by the magnetic scattering induced by net orbital magnetization associated with nonzero Berry curvature dipole. Our work paves the way to manipulate the Berry curvature in 2D materials by uniaxial strain.

## 2. Experiment and Method

WSe$_2$ flakes are exfoliated from bulk crystals onto silicon wafer with 285nm SiO$_2$. Monolayer WSe$_2$ can be identified by optical contrast and fluorescent microscopy. Then the monolayer WSe$_2$ is transferred onto the PMN-PT substrate by using a pick-up transfer technique[36]. Detailed topography of the transferred WSe$_2$ is imaged by the atomic force microscopy to resolve the possible ruptures and bubbles. Subsequently a uniform part of WSe$_2$ is selected to pattern into a standard Hall-bar geometry. Electrical contacts are made by e-beam lithography followed by e-beam evaporation of Ti (1nm)/Au (50nm) and lift-off in acetone. It is worth noting that the electrodes applying electric filed to the PMN-PT substrate is designed along the [001] orientation of the PMN-PT, by which the uniaxial strain can be produced. The strain in the monolayer WSe$_2$ introduced by the PMN-PT substrate is confirmed by the second-harmonic generation (SHG) spectroscopy. WSe$_2$ is heavily hole doped into metallic state by the ionic liquid technique and the low-temperature Ohmic contact is guaranteed.

## 3. Results and Discussions

The device structure and measurement scheme are depicted in Fig 1(a). The ionic gate voltage $V_G$ is fixed at -5 V and thus the holes in the valence bands at K/K' valleys dominates the transport properties[37]. Besides the gate volatge $V_G$, an extra volatge $V_{PMN-PT}$ is applied on the PMN-PT substrate after the ionic liquid is freezed below 180 K. The $V_{PMN-PT}$ produces an electric field $E_P$ between the electrodes along the [001] direction with the distance of 50 μm. The uniaxial strain is induced along $E_P$ direction into monolayer WSe$_2$ by the inverse piezoelectric effect of PMN-PT. To verify the strain in the WSe$_2$, the SHG spectroscopy is applied. Polarization resolved SHG reflects the lattice symmetry of the probed crystals and SHG intensity is susceptible to the mechanical deformation of the crystal lattice[38,39]. As shown in Fig 1(b), the SHG of the monolayer WSe$_2$ on the poled PMN-PT substrate is distorted from that of unstrained WSe$_2$ with the perfect D$_{3h}$ symmetry, establishing the induced strain in WSe$_2$ by PMN-PT. For the quantitative description of the strain level, we pattern and deposit a strain gauge made up of Ti/Au (1 nm/ 50 nm in thickness) on the same PMN-PT sustrate [Fig 1(c)] besides the deposition of the contatct metal. The strain measured from the strain gauge can roughly reflect the controllabe strain level that we apply to the WSe$_2$. Figure 1(d) shows the temperature dependence of the maximum tensile strain $\varepsilon$ of the strain gauge induced by the PMN-PT substrate at $E_P$ =18 kV/cm, which decreases from 0.14% to 0.02% upon decreasing temperature from 140 to 2 K. The decrease of the strain with decreasing temperature is consistent with the fact that the piezoelectric coefficient of PMN-PT decreases by a factor of 0.2 between 140 and 5 K[40].

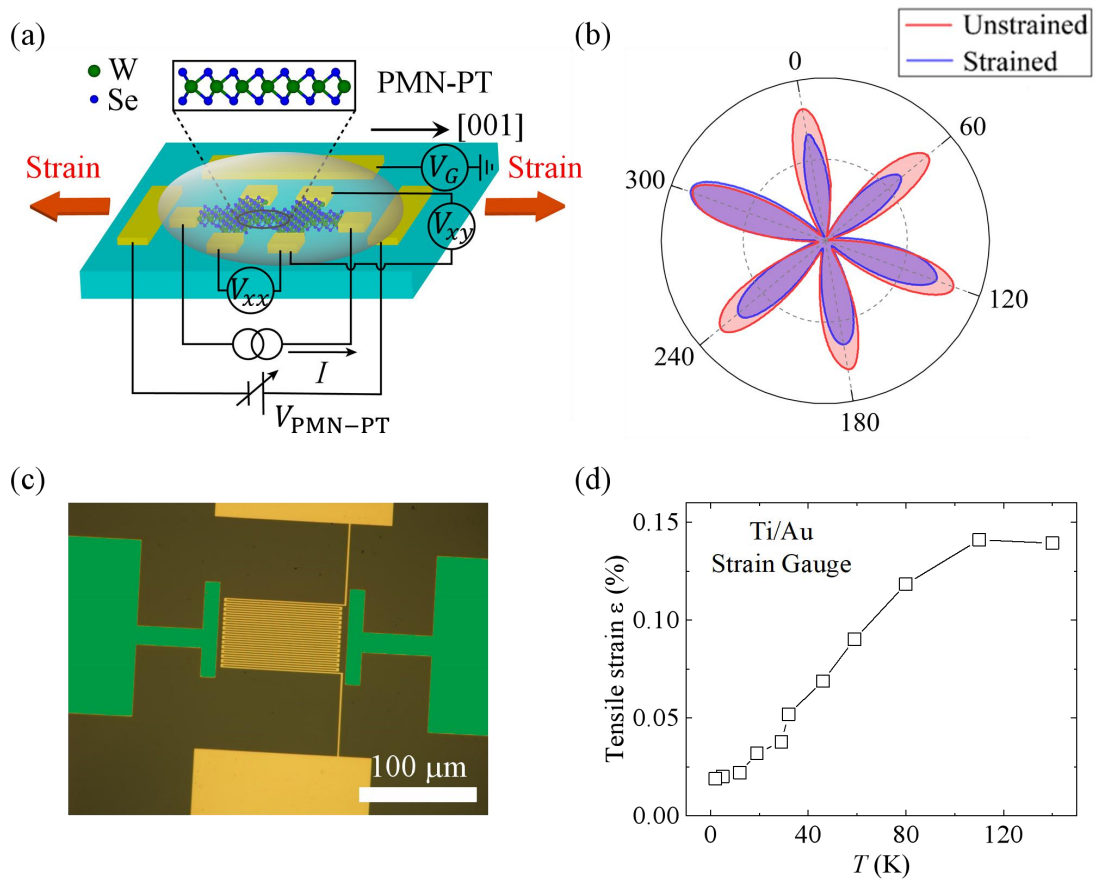

**Fig. 1**. (a) Schematics of ionic gated monolayer WSe$_2$ device. The electrodes applying

the $V_{PMN-PT}$ is designed along the [001] orientation of the PMN-PT substrate to ensure the uniaxial tensile strain. (b) Polarization resolved second harmonic generation intensity pattern. Red line: unstrained monolayer WSe$_2$. Blue line: strained monolayer WSe$_2$. (c) The optical microscope image of the Ti/Au strain gauge. (d) Temperature dependence of the maximum tensile strain produced in the PMN-PT substrate with a fixed polarized electric field $E_P$ =18 kV/cm, measured by the Ti/Au strain gauge.

Figure 2(a) shows the sheet resistance of monolayer WSe$_2$ as a function of $E_P$ measured at different temperatures. Before sweeping $E_P$, we first pole the PMN-PT along the [001] direction by applying $E_P$ =18 kV/cm for 15 minutes at 140 K to ahcieve a well aligned state and hence the initial strain in WSe$_2$ is fixed to be moderately tensile. Then we sweep $E_P$ with multiple cycling between $E_P$=±18 kV/cm until the strain-induced sheet resistance change stabalizes into an unipolar hysteretic loop, as shown in Fig 2(a). The emergence of the hysteretic behavior is due to the ferroelectricity nature of PMN-PT[41]. It is worth noting the coercive field of PMN-PT is strongly temperature dependent and it increases exponentially as the temperature decreases[42]. The coercive field at low temperatures is (~35 kV/cm at 40 K[42]) beyond the range of electric field that is achievable (±18 kV/cm) in this work. So when we backward sweep $E_P$ to -18 kV/cm, the piezoelectric polarization cannot be switched to the opposite direction at low tenperatures. In other words, the tensile strain varies monotonously without sign change by sweeping $E_P$ from positive to negative value. When we cool down the device from 140 to 2 K, the hysteresis grow weak and finally disappears, as shown in Fig 2(a), which is also attributed to the increased coercive field upon decreasing temperature.

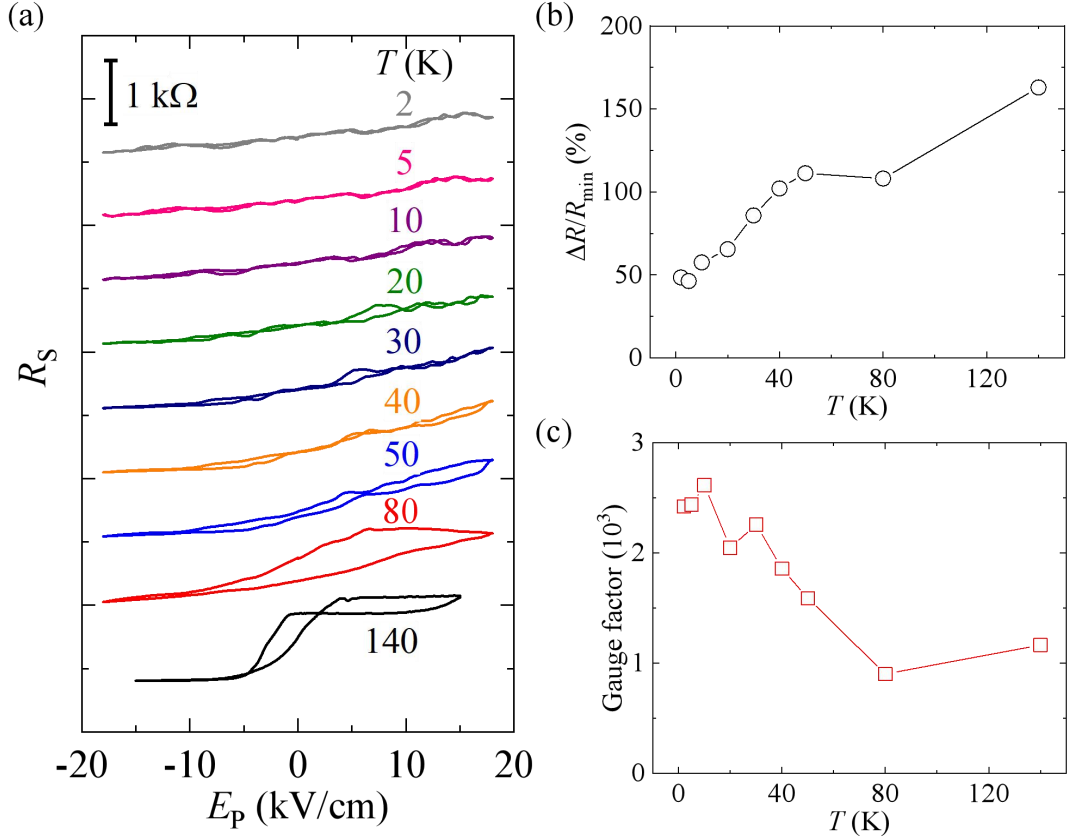

**Fig. 2.** (a) Sheet resistance of monolayer WSe$_2$ as a function of $E_P$ measured at temperatures from 140 to 2 K. The curves are shifted for clarity. (b) Ratio of percentual changes of the sheet resistance $\Delta R/R_{min} \times 100\%$ under the tensile strain and (c) gauge factor as a function of temperature.

To quantitativly investigate the strain dependence of resistance in WSe$_2$, we define the ratio of percentual changes of the sheet resistance under the maximum tensile strain ($E_P$ =18 kV/cm) by $\frac{\Delta R}{R_{min}} = \frac{R_{max} - R_{min}}{R_{min}} \times 100\%$. The $R_{max}$ and $R_{min}$ are defined as the resistance at $E_P$ =18 kV/cm and -18 kV/cm, respectively. As caluculated from Fig 2(a), $\frac{\Delta R}{R_{min}}$ monotonically decreases from 160% at 140 K to 48% at 2 K [Fig 2(b)]. The decreased resistance change is consistent with the suppressed strain level at low temperatures. We further calculate the so-called gauge factor (γ) of the WSe$_2$ by the formula $\gamma = \frac{\Delta R}{R_{min}} \cdot \frac{1}{\varepsilon}$. As summarized in Fig 2(c), the gauge factor is extremely high, approximately 1100 at 140 K, and increases to 2400 at 2 K, which is at least three orders of magnitude higher than the conventional silicon-based devices[27]. Interestingly, the gauge factor seems to be insensitivity to temperature above 80 K, suggesting such giant gauge factor may persisit up to room temperature.

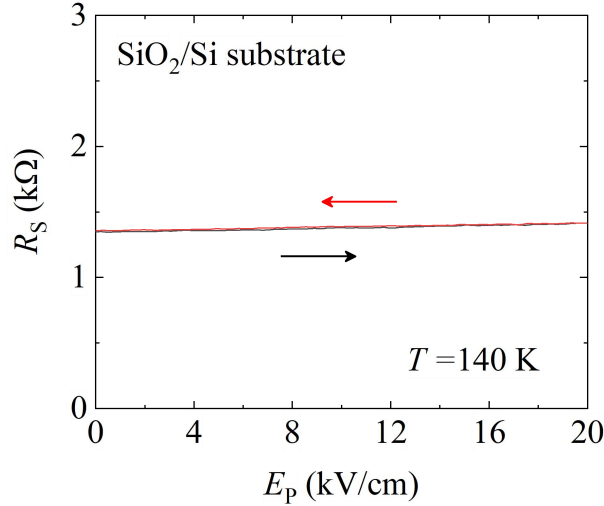

Fig. 3. Sheet resistance of monolayer WSe$_2$ as a function of $E_P$ measured at 140 K. The substrate is replaced by the non-piezoelectric SiO$_2$/Si substrate.

It is worth noting in addition to the strain effecct, extrinsic mechanisms, such as side gate effect and the influnece of $E_P$ on the ionic liquid, may also contribute to the changes of the sheet resistance of WSe$_2$. However, these extrinsic mechanisms can both be ruled out in this work. Although the side gate can effecctively tune the carrier density near the sample edges in a range of a few tens of nanometer in materials like graphene and InSb nanowire[43-45], the carrier density in the ionic gated WSe$_2$ (~$10^{13}$ cm$^{-2}$) is at least two orders of magnitude higher than these materials and hence the screening effect is prominent, which makes the side gating effect negligible in principle. We have performed a contrast experiment by using SiO$_2$/Si substrate to eliminate any possibility of the influnece of $E_P$ on the ionic liquid. We sweep $E_P$ as usual and observe negligible resistance change (less than 4%) and no hysterisis at 140 K, as shown in Figure 3.

We have noticed the strain-induced resistance change of TMDCs is previously ascribed to the bandgap engineering[20]. The gauge factor of monolayer MoS$_2$ has been reported by Kis *et al.* to be a relatively low negative value ~ -148 at 300 K [20]. However, the monolayer MoS$_2$ is undoped and semicoducting in that work. So the resistance change is attributed to the reduced bandgap indcued by tensile strain[14]. We stress monolayer WSe$_2$ used here is in the metallic phase with heavily hole-doping. Since the Fermi level is far from the band edge, the bandgap engineering can be safely excluded as the main cause of the giant gauge factor here.

To figure out the origin of the strong dependence of resistance on strain, the Hall carrier density and mobility as a function of $E_P$ at 2 K is calculated, as shown in Fig. 4(a). The Hall carrier density and mobility are calculated by the formula $n_H = \frac{1}{|e|R_H}$ and $\mu = \frac{R_H}{R_S}$, respectively, where the Hall coefficient $R_H$ is defined as $R_H = \frac{R_{xy}(+14\,\text{T}) - R_{xy}(-14\,\text{T})}{28\,\text{T}}$ and $e$ is the electron charge. The carrier density shows a rather small decrease ~10% between $E_P = \pm 18$ kV/cm, which may be ascribed to the work function change indcued by strain[19]. Importantly, the mobility is decreased about 35% between $E_P = \pm 18$ kV/cm, close to the resistance change ~48% at 2 K. Thus, we conclude the resistance change mainly originates from the decrease of mobility induced by strain.

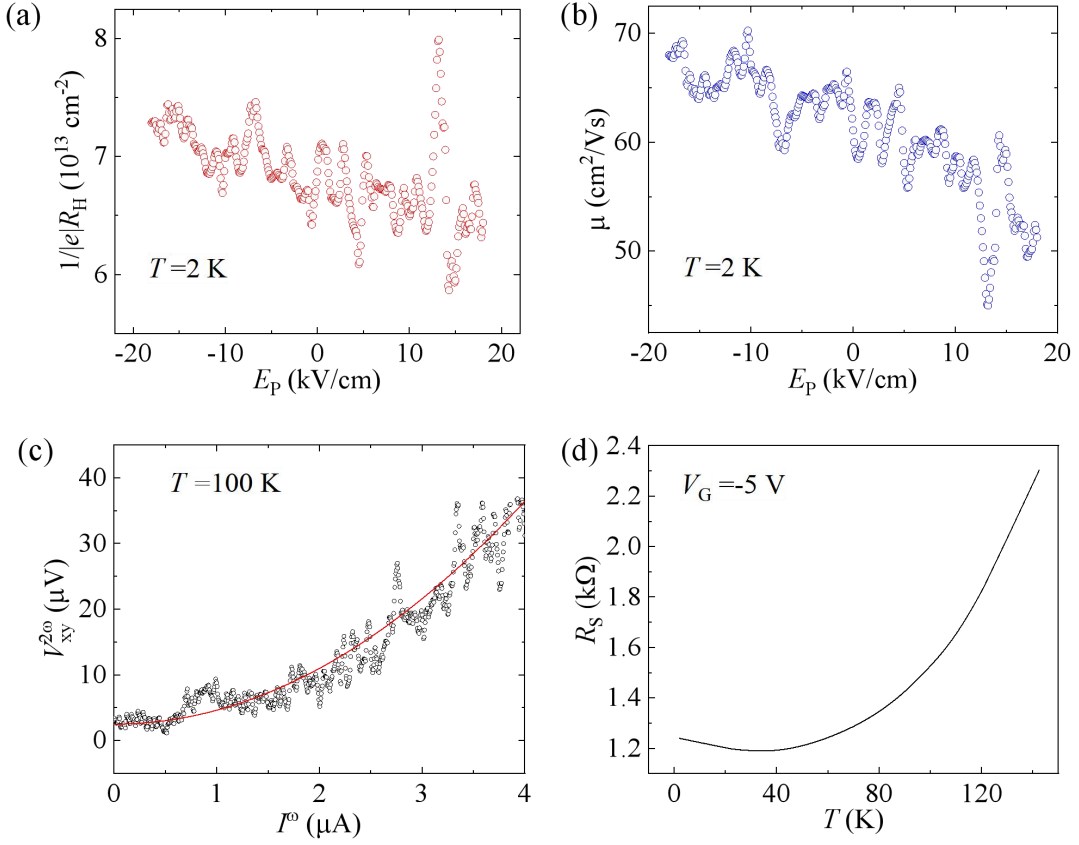

**Fig. 4.** (a) Hall carrier density and (b) mobility of $WSe_2$ as a function of $E_P$ at 2 K. (c) The second-harmonic Hall voltage $V_{xy}^{2\omega}$ versus $I^\omega$ at 100 K for a typical strained $WSe_2$ device. (d) The temperature dependent sheet resistance of monolayer $WSe_2$ with gate voltage $V_G$=-5 V.

Here we attribute the strain engineering of mobility to the strong magnetic scattering associated with nonzero Berry curvature dipole. For monolayer TMDCs, large Berry curvature emerges at K/K' valleys due to the spontaneous inversion symmetry breaking[28]. Berry curvature plays an important role in the topological transport, which would induce the transverse deflection of electron movement[46]. The presence of time-reversal symmetry in TMDCs leads to the opposite signs of Berry curvature in opposite valleys. Thus, driven by an in-plane electric field $E$, the electrons at opposite valleys will deflect along opposite directions and cancel with each other, leading to a charge neutral valley current, known as valley Hall effect[46]. In addition to Berry curvature, the dipole moment of Berry curvature, i.e., Berry curvature dipole, would also induce anomalou transport phenomena. However, the Berry curvature has a symmetric distributions in a single vallye constrained by the $C_{3v}$ symmetry, leading to vanishing Berry curvature dipole, which exactly describes the asymmetric distributions of Berry curvature[47]. Applying uniaxial strain can break the $C_{3v}$ symmetry in monolayer TMDCs, inducing nonzero Berry curvature dipole. The Berry curvature dipole $D$ is proposed to generate a net out-of-plane orbital magnetization $M \propto D \cdot E$ when an external electric field $E$ is also applied[47]. The emergent orbitial magnetization can induce a Hall current as a second-order response to $E$, that is, the nonlinear Hall effect[46], As shown in Fig. 4(c), the nonlinear Hall effect has been observed in a typical strained monolayer $WSe_2$. The emergent second-harmonic Hall voltage can be well fitted by parabolic curve [the red line presented in Fig. 4(c)], which is regarded as the evidence of nonlinear Hall effect[28]. Moreover, such orbital magnetization can also introduce additional magnetic scattering, thus reducing the mobility and longitudinal conductance. Obviously, a larger Berry curvature dipole means larger orbital magnetization, indcuing the stronger suppression of mobility. Note previous stuides on the nonlinear Hall effect in strained $WSe_2$ have shown that Berry curvature dipole could be significantly enhanced by increasing the tensile strain[28,47]. Thus, here the measured giant gauge factor can be ascribed to the strain induced additional magnetic scattering associated with a nonzero Berry curvature dipole. Furthermore, as shown in Fig. 4(d), the temperature dependent resistance of $WSe_2$ shows a metallic behavior at high temperature, while it increases upon decreasing temperature below 40 K. Such temperature-dependent behavior is also consistent with the existence of magnetic moments, which can lead to the Kondo effect and a minimum resistance.

## 4. Conclusions

We have demonstrated a strong dependence of resistance on uniaxial strain in monolayer $WSe_2$ at various temperatures. The gauge factor value can reach as large as 2400, demonstrating great application potential for the next-generation ultra-sensitive sensors. We attribute this strain-dependent resistance to the strong magnetic scattering associated with nonzero Berry curvature dipole induced by strain. Our work indicates that uniaxial strain can effectively modulate the Berry curvature via changing the symmetry and band structure in 2D vdW materials, providing an effective strategy to explore the strain related effects on numerous 2D materials.

Acknowledgements: Supported by the National Key Research and Development Program of China (Grant No. 2018YFA0703703), and the National Natural Science Foundation of China (Grant Nos. 91964201, 61825401, and 11774004).